# Optical properties of BAlN and BGaN for applications in lattice-matched UV optical structures

**Feras AlQatari[1], Muhammad Sajjad[2], Ronghui Lin[1],**

**Kuang-Hui Li[1], Udo Schwingenschlögl[2], and Xiaohang Li[1,*]**

*[1] Advanced Semiconductor Laboratory, King Abdullah University of Science and Technology (KAUST), Thuwal 23955-6900, Saudi Arabia*

*[2] Physical Science and Engineering Division, King Abdullah University of Science and Technology (KAUST), Thuwal 23955-6900, Saudi Arabia*

*Corresponding author: xiaohang.li@kaust.edu.sa

## Abstract

The optical properties of BAlN and BGaN ternary alloys are investigated using first-principle calculation. Hybrid density functional theory is applied to determine the refractive indices of different alloys. A peculiar non-linear behavior of the static refractive index as a function of boron composition is found. The results of this calculation are interpolated to generate a three dimensional dataset, which could be used for designing a myriad of strained and strain-free optoelectronic and photonic devices. This is then used to find a lattice-matched heterostructure optimized for DBR applications ($B_{0.108}Ga_{0.892}N$/AlN). A DBR design with 25 pairs at a wavelength of 375 nm is found to have peak reflectivity of 99.8% and a bandwidth of 26 nm.

## Introduction

Wurtzite III-nitride semiconductor materials have many technically important properties for applications in optical and electronic devices[1]. The III-nitride based optical devices can operate over a wide spectrum of wavelengths from infrared (IR) to ultraviolet (UV)[2] due to highly tunable bandgaps from 0.7 to 6 eV[3]. The interest in developing UV devices such as UV light-emitting diodes and UV lasers is steadily rising for vital applications such as sterilization, data storage, biochemical sensing, and communication[4-13].

Optical structures such as distributed Bragg reflectors (DBRs) are essential in UV surface-emitting lasers for cavity formation and UV LEDs for higher light extraction efficiency[14-16]. DBRs are dielectric or semiconducting superlattices capable of achieving near 100% reflectivity. The formation of cracks due to large strain and lattice mismatch can greatly compromise the performance of DBRs[27]. Thus, both lattice matching and large refractive index contrast are preferable especially for epitaxial DBRs. Conventionally, AlGaN/AlGaN superlattices have been employed for UV applications where the higher index contrast leads to higher the lattice mismatch[53,55]. In theory, InAlN would be ideal for lattice matching with AlGaN alloys, though it is challenging to grow such alloys with high quality due to large lattice difference, phase separation, and temperature incompatibility between InN (~600°C) and AlN (≥1100°C)[17,18]. This leads to relatively complicated growth methods for the realization of near lattice-matched AlGaN/InAlN DBR with good performance[19].

Additionally, extensive research activities have been performed on metalenses, which allow for precise control of optical wavefronts with the potential to reduce the size and weight of bulky refractive lenses leading to countless opportunities such as space optical communication, portable light weight imaging systems and many more[20,21,22]. Knowing the optical properties and bandgaps of III-nitride materials at various lattice constants is necessary for designing these structures.

Recently, the search for more materials in the III-nitride system has inspired a number of studies of boron-containing III-nitrides including BAlN and BGaN[23-30]. Alloying III-nitrides with boron can reduce their lattice parameters giving a new option for strain engineering and lattice matching[31]. In addition, the incorporation of boron could lead to more alloys with large bandgaps adding more options for device engineering.

The refractive index is important for the design and simulation of optoelectronic devices in the UV range. It has been experimentally shown that a small incorporation of boron into GaN and AlN can cause a significant change in the refractive index of materials[23,24].

Recently, the full range of compositions for BGaN alloys has been investigated by Said et al using local density approximation (LDA)[26] of density functional theory (DFT). However, local functionals such as LDA largely underestimate the bandgaps, redshift absorption spectra and miss some excitonic features[32]. Additionally, the lattice parameters in their study show a significant underestimation compared with experimental data. This indicates that the structures may be strained possibly as a result of non-convergence in the structure relaxation step.

In this work, we employed the HSE functional which gives more accurate results for optical properties of wide bandgap materials. Additionally, investigating three alloys systems, AlGaN, BAlN and AlGaN, using the same method in the one study gives more reliable trends, and closer estimations of differences in refractive indices. Our investigation of the three alloy systems was done over the entire compositional range. The results of this calculation were then interpolated, giving a continuous image under all possible compositions. Finally, the results of interpolation were used to find an optimum in terms of DBR performance using finite element simulations.

## Computational method

All reported calculations were carried using the software VASP (Vienna Ab initio Simulation Package)[33,34]. Before calculating the refractive indices the structure was optimized using DFT general gradient approximation (GGA-PBEsol) exchange-correlation[35]. The energy cutoff was set to 520 eV for the plane-wave basis set. The structure optimization was performed on primitive cells for the binary systems and on 16-atom supercells for the ternary systems with chalchopyrite (CH) and luzonitelike (LZ) structures for the 50%, and 25% / 75% alloys, respectively, as reported in our previous study[28]. For 12.5% we used the same CH structure as 25%, but with the replacement of a B atom with Al or Ga. The reason the calculation is performed for 12.5% is to see a clearer image at low B composition where most experimental work exists[27,29-31]. All structures were relaxed to Hellman-Feynman force less than 0.02 $eVÅ^{-1}$. The Gamma centered k-point mesh was set to $6 \times 6 \times 6$ for the structural optimizations.

To calculate the optical properties, the hybrid functional of Heyd, Scuseria, and Ernzerhof (HSE) was used[36]. Similar to structure optimization, 520 eV was used as the energy cutoff for the plane-wave basis set. The k-meshes of $8 \times 8 \times 8$ and $6 \times 6 \times 6$ were used for the binary and ternary systems, respectively. The same parameters were used for the three systems AlGaN, BAlN, and BGaN for the structure optimization step as well as for the optical properties calculation step.

The results of this calculation were then plotted in three dimensions for better visualization. The generated line plots were 2D-interpolated using the Modified Akima cubic Hermite method. This interpolation method was chosen due to its lower variations in the predicted variable than when using the spline method. Additionally, it has less sharp edges when compared to linear interpolation. This, we believe, would give a more sensible prediction of refractive index curves with a high compositional resolution.

The resulting interpolated surface was used to find an optimum design for a strain-free, high-reflection and high-bandwidth DBR. The performance of this DBR was simulated using the software Lumerical FDTD Solutions[37]. An assumption of no optical absorption was made, which is reasonable for a semiconductor below its bandgap energy.

## Results and discussion

**Structure** First, we confirm that the optimized wurtzite structures have lattice parameters within a reasonable range. Table 1 shows our calculated lattice parameters, where binary structures show good agreement with experimental data. In addition, ternary alloys show good agreement with other theoretical calculations as is discussed in our previous study[28].

Table 1: Lattice parameters of different wurtzite III-nitride alloys.

| ***Alloy*** | | ***a(Å)*** | ***c(Å)*** |
|---|---|---|---|
| *AlN* | **This work** | **3.113** | **4.981** |
| | Experiment[38] | 3.11 | 4.98 |
| $B_{0.125}Al_{0.875}N$ | This work | 3.054 | 4.869 |
| $B_{0.25}Al_{0.75}N$ | This work | 2.984 | 4.783 |
| $B_{0.5}Al_{0.5}N$ | This work | 2.876 | 4.556 |
| $B_{0.75}Al_{0.75}N$ | This work | 2.713 | 4.372 |
| *GaN* | **This work** | **3.182** | **5.180** |
| | Experiment[38] | 3.180 | 5.166 |
| $B_{0.125}Ga_{0.875}N$ | This work | 3.107 | 5.049 |
| $B_{0.25}Ga_{0.75}N$ | This work | 3.020 | 4.943 |
| $B_{0.5}Ga_{0.5}N$ | This work | 2.898 | 4.673 |
| $B_{0.75}Ga_{0.25}N$ | This work | 2.722 | 4.431 |
| *BN* | **This work** | **2.543** | **4.205** |
| | Experiment[38] | 2.55 | 4.21 |

**Absorption and refractive indices** The focus of this study is mainly on the optical properties below the bandgap, where absorption is typically assumed to be too small. The calculated values are reported in Figure 1. All reported values are for light polarization perpendicular to the *c*-axis. Both absorption and refractive indices are calculated using the complex dielectric function as described in the following definition[39]:

$$n = \frac{1}{\sqrt{2}}\sqrt{\sqrt{{\varepsilon_1}^2 + {\varepsilon_2}^2} + \varepsilon_1},\ \kappa = \frac{1}{\sqrt{2}}\sqrt{\sqrt{{\varepsilon_1}^2 + {\varepsilon_2}^2} - \varepsilon_1} \tag{1}$$

$n$ is the refractive index, $\kappa$ is the absorption index, and $\varepsilon_1$ and $\varepsilon_2$ are the real and imaginary parts of the dielectric function, respectively.

As $\varepsilon_2$ approaches zero, $\kappa$ approaches zero as well, and $n$ approaches the square root of $\varepsilon_1$. This usually occurs below the bandgap (approaching 0 eV). This assumption is used in later sections of this study.

Before discussing the results for BAlN and BGaN alloys, we first confirm our model by comparing the refractive indices of the AlGaN alloys with the reported experimental results as shown in Figure 1 (a&b). The calculated values of AlGaN are consistent with experimental data with a small underestimation proportional to gallium composition[40,41]. Figure 1 shows the calculated refractive indices of all three alloys.

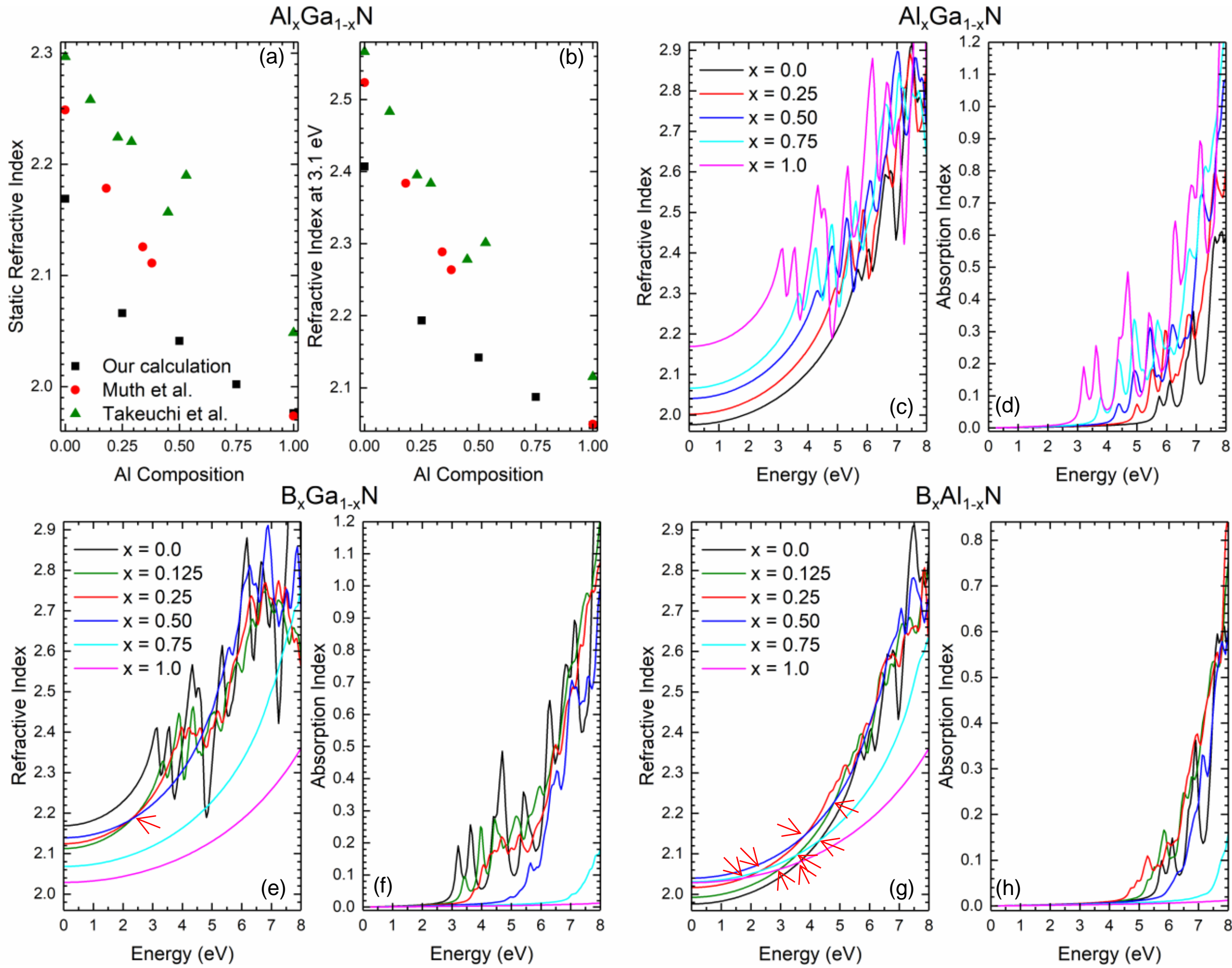

*Figure 1: Comparison of the refractive index of $Al_xGa_{1-x}N$ as calculated by us and as experimentally reported by Muth et al.*[41] *and Takeuchi et al.*[40] *(a) The static refractive index (at 0 eV). (b) The refractive index near $E_g^{GaN}$. Refractive and absorption indices of $Al_xGa_{1-x}N$ (c & d), $B_xGa_{1-x}N$ (e & f), $B_xAl_{1-x}N$ (g & h) as a function of photon energy. Red arrows in (e&g) point at the cross points of index plots below $E_g$.*

An interesting behavior of the refractive index appears as the boron content is increased from zero. For both BAlN and BGaN, the refractive index follows a non-linear trend. This can be visualized by the static refractive index (the index at 0 eV) in Figure 2. The reason we focus on the static refractive index is that it captures the effect of alloying without having to consider the energy of any incoming photons (Equations 2&3). Typically, as the relaxed lattice parameter is increased by changing the composition, the bandgap is decreased and the refractive index is increased. This is obeyed by the AlGaN alloys as well as most III-nitride materials[42]. Other studies of boron III-V semiconductors—such as BAlAs, BAlSb and BGaAs—have found a similar non-linear behavior as in our study when boron content is increased[43-45]. This behavior has been attributed to the large difference in atomic size and electronegativity. In addition, unlike other group III elements, boron lacks inner p states, contributing to the unusual properties of boron alloys.

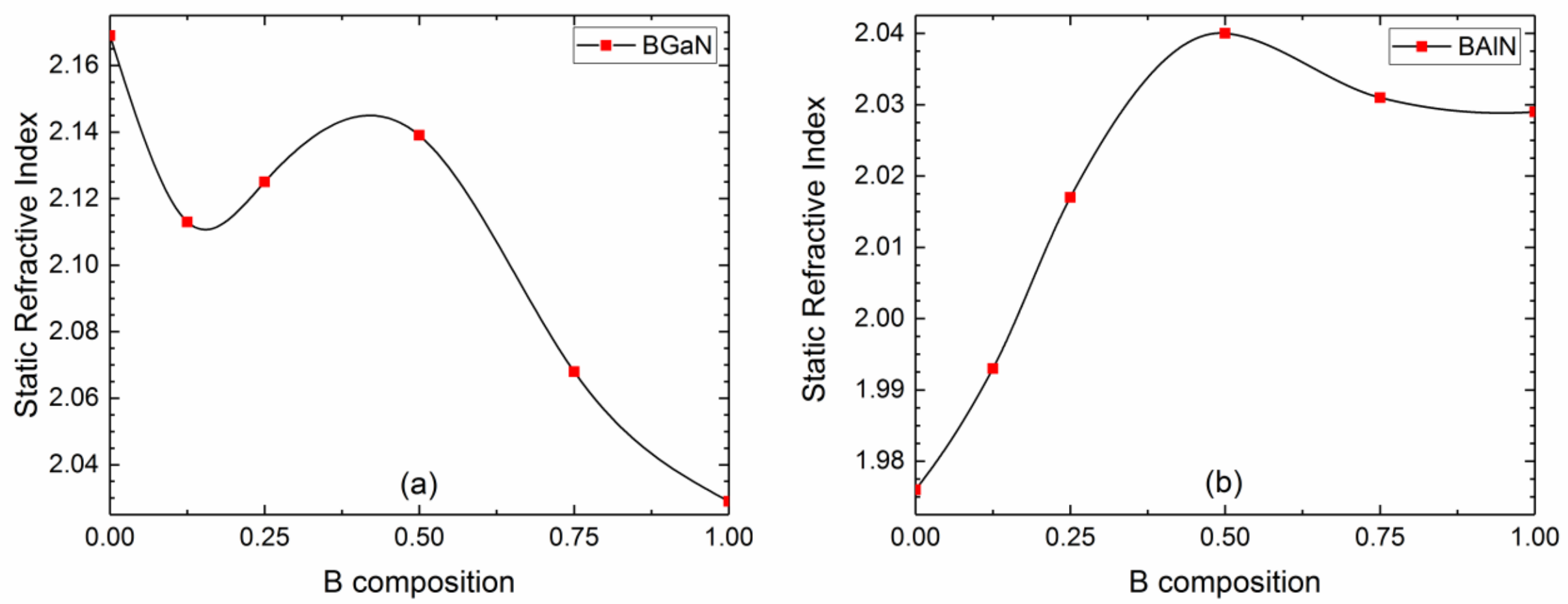

*Figure 2: Static refractive indices of BGaN (a) and BAlN (b) as a function of boron composition.*

Additionally, we can observe by going back to Figure 1 (e&g) that at certain compositions the index plots cross. This occurs for low boron content BGaN alloys ($x \leq 0.5$). In BAlN, this effect is clearer and occurs over many alloy concentrations. A similar crossing of index curves is observed in BAlSb as shown by Benchehima et al[44].

Several empirical models have been developed to predict the index of materials[42,46-48]; however, in this study, we mainly examine the classical model of light propagation inside materials for a clue to why the index behaves this way. This model is chosen due to its simplicity and because it is derived from first principles.

$$n(\omega) = 1 + \frac{Nq_e^2}{2\epsilon_0 m}\sum_k \frac{f_k}{\omega_k^2 - \omega^2 - i\gamma_k \omega} \quad (2)$$

$N$ is the electron density, $q_e$ is the electron charge, $\epsilon_0$ is the vacuum permittivity constant, $m$ is electron mass, $\omega_k$ is the frequency for each electron transition, $\omega$ is the frequency of light, $\gamma_k$ is the damping rate for each electron, and $f_k$ is oscillator strength of each electron.

Equation 2 predicts a jump in the refractive index near $\omega_k$ for each mode k. These oscillation modes are associated with electron transitions in the band structure[49]. The first electron transition occurs near the direct bandgap.

At 0 eV, Equation 2 becomes:

$$n(0) = 1 + \frac{Nq_e^2}{2\epsilon_0 m}\sum_k \frac{f_k}{\omega_k^2} \quad (3)$$

Equation 3 indicates that the static refractive index is inversely proportional to $\omega_k^2$ for each mode k. The bowing in the electron transition energy associated with each mode may be different. This approach is similar to work done by Shen et al. different bowing parameters are assigned for different electron transitions[50]. We hypothesize that the non-linearity in the static refractive index may be the result of the different behavior of different electron transitions. A similar consideration was made by Linnik et al. to develop a model describing cubic zinc-blende III-V semiconductors where they used an interband transition contributions (ITC) model to describe the dielectric function[49]. This is a more in depth explanation when compared with simply attributing the behavior to atomic size or electronegativity though both views may be related. It is interesting to note that Takeuchi et al. applies an interband transitions model to the AlGaN system, though no bowing is assumed[40].

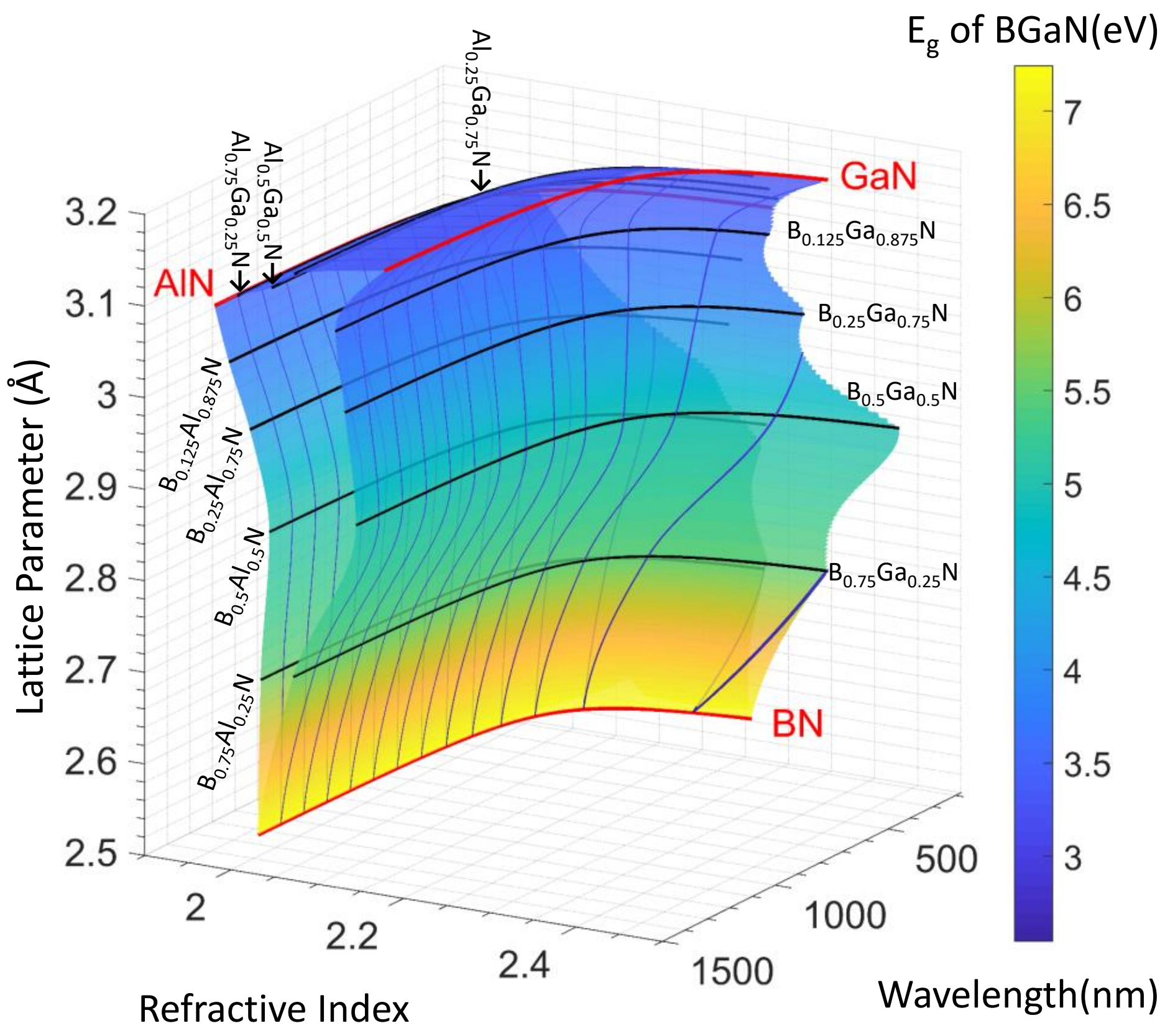

*Figure 3: Three dimensional visualization of the index (1.9-2.5) of each material (lattice: 2.5-3.2 Å) in the 175-1500 nm range of wavelengths. The color bar represents the bandgap of the lower bandgap material BGaN at each lattice parameter. The surface is cutoff at the bandgap of BGaN as given by the self-consistent field.*

The data presented in Figure 1 combined with Table 1 can be presented more concisely and more elegantly with the help of interpolation. Figure 3 shows a three-dimensional visualization of the refractive indices of all compositions of the studied alloys as a function of wavelength and lattice parameter. The black horizontal curves represent the indices at the calculated compositions as shown in Figure 1 (c,e&g). The surfaces connecting the lines are interpolated to show the expected indices at all other compositions. The vertical blue lines represent borders of 100 nm thick slices of the 3D object, added to show depth.

It is important to note that the data in Figure 3 is cutoff below the bandgap of BGaN shown in the color bar and with experimental values. The cutoff value is chosen because it is where the optical transitions occur in the optical data, leading to jumps in index values at energies lower than found experimentally. The color bar value is important because it represents the maximum operation frequency of a lattice-matched material pair in a device before absorption becomes an issue. The bandgap of BGaN is used for the color bar since it has the smaller bandgap of any pair at any lattice parameter value. Similar to this study, the color bar values are calculated using hybrid density functional theory and are reported in a separate study[51].

Using this figure we can find a virtually infinite number of lattice matched material pairs. This is very important for designing crack-free photonic and optoelectronic devices. To use our results more easily, all data in this work will be available on the website nitrideoptics.com.

The difference in refractive index is a parameter that allows us to estimate the performance of different optoelectronic and photonic devices such as DBRs.

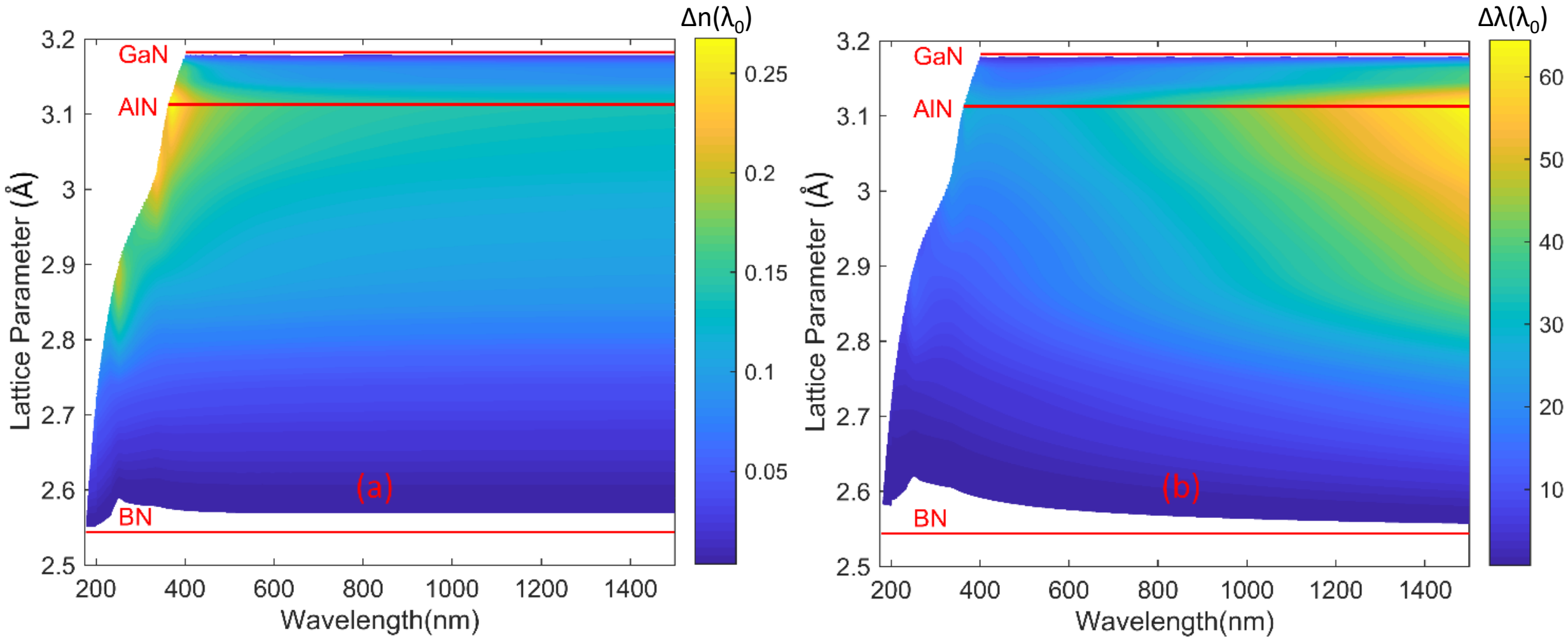

(a) shows the difference in index at each wavelength. This quantity can be directly extracted from Figure 3. To understand Figure 4 (a) we can think of Figure 3 as a physical 3D object. Using this image, the color bar can be thought of as the thickness of Figure 3 along the axis of the index. Figure 4 (b) shows the expected bandwidth of a DBR as defined by Equation 4 at each wavelength[52]:

$$\Delta\lambda = \frac{\pi}{2}\lambda_0\left[\frac{1}{arccos(\rho)} - \frac{1}{arccos(-\rho)}\right], where\ \rho = \frac{n_H - n_L}{n_H + n_L} \quad (4)$$

$\lambda_0$ is the design wavelength, $\rho$ is reflection coefficient, $n_H$ and $n_L$ are high and low refractive indices, respectively.

*Table 2: Camparison of performance of different DBR structures with similar peak wavelengths. Lattice mismatch is calculated based on Vegard's law. Bandwidth is taken at 90% of peak performance.*

| ***Result Type*** | ***Structure*** | ***Lattice Mismatch*** | ***$\lambda_0$ (nm)*** | ***Pairs*** | ***Reflectivity*** | ***$\Delta\lambda$ (nm)*** |
|---|---|---|---|---|---|---|
| ***This work*** | **$B_{0.108}Ga_{0.892}N/AlN$** | **0%** | **375** | **25** | **0.998** | **26** |
| *Experimental*[19] | $Al_{0.83}In_{0.17}N/Al_{0.2}Ga_{0.8}N$ | 0.511% | 368 | 45 | 0.993 | 13 |
| *Experimental*[53] | $Al_{0.12}Ga_{0.88}N/GaN$ | 0.265% | 368 | 40 | 0.916 | 4.4 |
| *Simulation* | | | 369 | 40 | 0.972 | 5.6 |
| *Experimental*[54] | $Al_{0.18}Ga_{0.82}N/Al_{0.8}Ga_{0.2}N$ | 1.37% | 347 | 20 | 0.93 | 18 |
| *Experimental* | | | 347 | 25 | 0.99 | 22 |
| *Experimental*[55] | $Al_{0.2}Ga_{0.8}N/GaN$ | 0.436% | 380 | 60 | 0.99 | 8 |

*Figure 4: Visualization of the difference in refractive index ($\Delta n$) (a) and the predicted theoretical DBR bandwidth ($\Delta\lambda$) (b) as functions of the design wavelength of the DBR ($\lambda_0$). The color of each point represents the value corresponding to a strain-free material pair at the indicated lattice parameter and wavelength. Red lines indicate the lattice parameters of binary materials.*

We can see clearly an optimum behavior occurring at 3.113 Å by looking at Figure 4. This represents the material pair $B_{0.108}Ga_{0.892}N/AlN$ where the boron composition is calculated using Vegard's law. We use this material pair to simulate a DBR at the wavelength 375 nm in the UVA range. This wavelength is chosen due to our interest in developing the first electrically pumped UVA VCSEL.

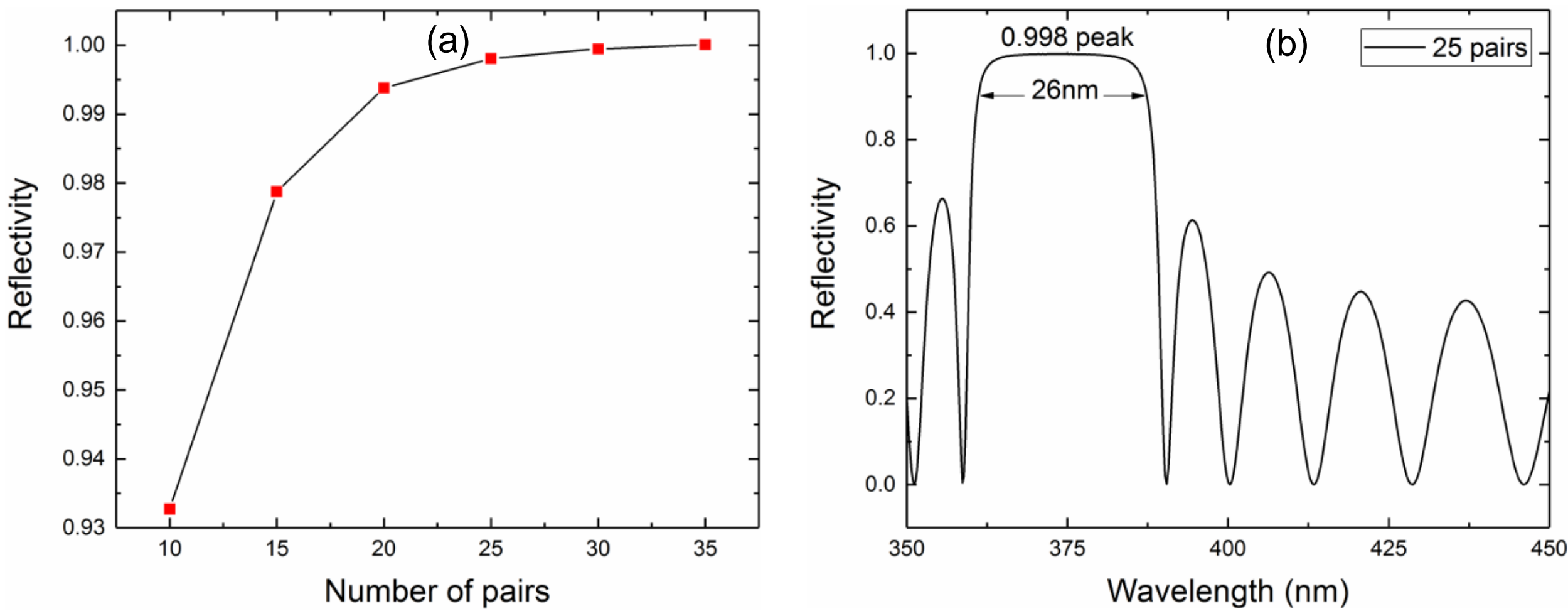

*Figure 5: DBR peak reflectivity as a function of the number of pairs (a) and Reflectivity as a function of wavelength for a 25 pair DBR (b). Both figures are simulated for the optimized material pair* $B_{0.108}Ga_{0.892}N/AlN$.

First we start by sweeping different numbers of pairs to see change in peak reflectivity at 375 nm. We can see in Figure 5 (a) that the DBR performance exceeds 99% reflectivity near 20 pairs and plateaus near 25 pairs. Reflectivity reaches 100% at 35 pairs in this simulation. The reflectivity spectrum of the 25-pair device is shown in Figure 5 (b), where peak reflectivity is 99.8%. Bandwidth measured at 90% performance is 26 nm. This high reflectivity and wide bandwidth is ideal for optimal UVA VCSEL performance[53]. Table 2 shows a comparison of this DBR structure with similar structures that are designed for UVA VCSEL applications.

## Conclusion

We have investigated the refractive index of BAlN and BGaN alloys using hybrid density functional theory. A non-linear behavior was found in the static refractive indices of different alloys. The bowing effect in the bandgaps of these alloys may provide an explanation. A more in-depth study of the electronic band structure is needed to prove such hypothesis. However, regardless of the physical origin of this behavior, the interpolated dataset could be used in designing a myriad of strain-free photonic and optoelectronic devices. We have demonstrated an optimized DBR at 375 nm having a peak reflectivity of 99.8% and a bandwidth of 26 nm.